\documentclass[12pt]{article}
\usepackage{bbm}
\usepackage{epsfig}
\usepackage{array}
\usepackage{float} 
\usepackage{amsmath} 
\usepackage{amssymb}
\usepackage{amsfonts}

\usepackage{a4}
\usepackage{epsfig}

\usepackage{a4wide}
\usepackage{wasysym}

\def\ra{\rightarrow}
\def\be{\begin{equation}}
\def\ee{\end{equation}}
\def\gs{\mathrel{
   \rlap{\raise 0.511ex \hbox{$>$}}{\lower 0.511ex \hbox{$\sim$}}}}
\def\ls{\mathrel{
   \rlap{\raise 0.511ex \hbox{$<$}}{\lower 0.511ex \hbox{$\sim$}}}}

\newcommand{\ba}{\begin{array}{c}}
\newcommand{\baz}{\begin{array}{cc}}
\newcommand{\bad}{\begin{array}{ccc}}
\newcommand{\bav}{\begin{array}{cccc}}
\newcommand{\baf}{\begin{array}{ccccc}}
\newcommand{\bas}{\begin{array}{ccccccc}}
\newcommand{\bea}{\begin{equation} \begin{array}{c}}
\newcommand{\eea}{ \end{array} \end{equation}}
\newcommand{\ea}{\end{array}}
\newcommand{\D}{\displaystyle}
\newcommand{\dms}{\mbox{$\Delta m^2_{\odot}$}}
\newcommand{\dma}{\mbox{$\Delta m^2_{\rm A}$}}

\newcommand{\eV}{\mbox{ eV}}

\newcommand{\sss}{\sin^2 \theta_{12}}

\begin{document}

\title{\vspace{-1cm}
\hfill {\small arXiv:0803.2417 [hep-ph]} 
\vskip 0.6cm
\bf \Large 
A Yukawa coupling parameterization for \mbox{type I\,+\,II}   
seesaw formula and applications to lepton flavor violation and 
leptogenesis}

\author{
Evgeny Kh.~Akhmedov$^{a,b}$\thanks{email: \tt 
akhmedov@mpi-hd.mpg.de} 
\,~and ~Werner Rodejohann$^{a}$\thanks{email: \tt 
werner.rodejohann@mpi-hd.mpg.de} 
\\\\
{\normalsize \it $^{a}$Max--Planck--Institut f\"ur Kernphysik,}\\
{\normalsize \it Postfach 103980, D--69029 Heidelberg, 
\vspace*{0.4cm}%
Germany}\\ 
{\normalsize \it $^{b}$National Research Center 
Kurchatov Institute,}\\ 
{\normalsize \it 123182 Moscow, Russia}
}
\date{}
\maketitle
\thispagestyle{empty}
\vspace{-0.8cm}
\begin{abstract}
\vspace*{0.2cm}
\noindent 
In the type I + II seesaw formula the mass matrix of light neutrinos $m_\nu$
receives contributions from the exchanges of both heavy Majorana neutrinos and 
$SU(2)_L$-triplet Higgs bosons. We propose a new parameterization for the
Dirac-type Yukawa coupling matrix of neutrinos in this case, which 
generalizes the well known Casas-Ibarra parameterization to type I + II 
seesaw and is useful when the triplet term in $m_\nu$ is known. Neutrino 
masses and mixing, lepton flavor violation in decays like $\mu \ra e  \gamma$ within mSUGRA models and leptogenesis can then be studied within 
this framework. We illustrate the usefulness of our new parameterization using 
a number of simple examples. 
\end{abstract}

\newpage

\section{\label{sec:intro}Introduction}
The seesaw mechanism \cite{I,II} provides a very natural 
and attractive explanation of the smallness of 
neutrino mass \cite{reviews} as being due 
to exchanges of heavy particles. In the most commonly 
considered type I seesaw, these are heavy sterile 
(electroweak-singlet) Majorana neutrinos 
\cite{I}; another well studied case is type II seesaw, 
where the small neutrino mass is generated by the 
induced vacuum expectation value (VEV)  
of an $SU(2)_L$-triplet Higgs boson \cite{II}.  
In both cases, the light 
neutrinos are Majorana particles with the effective mass 
matrix $m_\nu$, which in the basis where the mass 
matrix of the charged leptons is diagonal 
and real, is diagonalized according to   
\be 
\label{eq:mnu}
m_\nu = U^\ast \, m_\nu^{\rm diag} \, U^\dagger\,, \qquad
m_\nu^{\rm diag} = {\rm diag}(m_1, m_2, m_3)\,. 
\ee
Here $U$ is the leptonic mixing matrix, which depends on 
three mixing angles, 
one Dirac-type and two Majorana-type CP-violating phases. 
Although both sterile neutrinos and Higgs triplets can be 
freely added 
to the standard model, they are most natural in its partially 
unified or grand unified extensions, such as left-right 
symmetric models or $SO(10)$ 
grand unified theories (GUTs), where both type I and 
type II contributions 
to the neutrino mass are typically present. In that case 
the neutrino mass 
matrix is a sum of two terms:%
\footnote{Note that sometimes in the literature the 
mechanism leading to 
the entire Eq.~(\ref{eq:II}) (rather than only to the triplet 
contributions to $m_\nu$) is called type II seesaw.}
\be \label{eq:II}
m_\nu = m_\nu^{II} + m_\nu^{I} 
= v_L \, f_L - \frac{v_u^2}{v_R} \, Y_D \, f_R^{-1} \, Y_D^T\, .
\ee
Here the first term is the $SU(2)_L$-triplet Higgs contribution 
with $v_L$ the VEV of the triplet and $f_L$ the triplet 
Yukawa coupling matrix. The 
triplet VEV $v_L\simeq \mu \, v_u^2 /M_\Delta^2$, where $\mu$ is 
the trilinear Higgs coupling, 
$v_u$ is the VEV of the up-type Higgs doublet $H_u$,  
and $M_\Delta$ is the mass of the triplet. The second 
term in (\ref{eq:II}) is the conventional type I seesaw term, 
in which $v_u \, Y_D$ is the Dirac mass 
matrix $m_D$. Having in mind extensions of the 
standard model, we have written the Majorana mass matrix 
of heavy neutrinos $M_R$ as $v_R \, f_R$, with $f_R$ 
being the relevant coupling matrix. In particular, in 
left-right symmetric gauge theories $f_R$ is the Yukawa coupling matrix 
of an $SU(2)_R$-triplet Higgs and $v_R$ is its VEV, which is related to the 
VEV of the $SU(2)_L$-triplet via $v_L\,v_R \propto v_u^2$. Regardless of the 
variant of the seesaw mechanism and barring unnaturally small Yukawa couplings 
or strong cancellations, the typical mass scale of the neutrino mass generation 
($M_\Delta$ or $v_R$ or both) exceeds $10^9$ GeV, which is way beyond the 
reach of direct experimental tests. Hence, the seesaw mechanism of neutrino 
mass generation can only be probed indirectly.

One way of indirectly probing the seesaw is provided 
by cosmology, 
where the observed baryon number of the universe can 
be generated through  
the baryogenesis via leptogenesis mechanism \cite{leptog,D}. 
Leptogenesis 
can work successfully within both type I \cite{Ilepto,CRV} and 
type II \cite{leptoIIpure} seesaw scenarios,%
\footnote{In the case of pure type II seesaw more than 
one Higgs triplet 
is necessary for leptogenesis to work.}
as well as in the combined \mbox{type I\,+\,II} seesaw 
\cite{HamSen} (for earlier works see \cite{UO,JPR0,JPR1}), 
where the neutrino mass 
matrix is given by Eq.~(\ref{eq:II}). 
Another possibility of testing the seesaw is through 
lepton flavor violation 
(LFV) within supersymmetric theories, which has also 
been discussed in 
both type I \cite{ILFV} and type II \cite{IILFV,Rossi,LFVIIana} 
seesaw frameworks. Here 
LFV is induced by off-diagonal entries in the slepton 
mass matrices, which 
can be generated radiatively. In many well motivated scenarios, the size 
of these entries depends on the seesaw parameters. 
Leptogenesis and LFV depend on combinations of the Yukawa coupling 
matrices that are different from those entering into the seesaw 
formula (\ref{eq:II}), and this can be used -- at 
least in principle -- to reconstruct the seesaw parameters.

An important issue in these approaches is that the number of high energy 
parameters, i.e.~of those contained in $m_\nu^{II}$, 
$m_D$ and $M_R$, 
exceeds the number of low energy parameters contained in $m_\nu$, 
simply 
because the heavy degrees of freedom are integrated out at 
low energies. An 
exception is the case of type II dominance, when $m_\nu$ 
coincides (or 
approximately coincides) with $m_\nu^{II}$.  In the 
general case, however, and 
without a specific model at hand, one can only 
parameterize the unknown high 
energy quantities. Within the pure type I seesaw, 
one such parameterization, 
which proved to be especially useful and convenient, 
was suggested by Casas 
and Ibarra \cite{CI}. This is the parameterization of 
the Dirac-type Yukawa coupling matrix $Y_D$ in 
which it is written as  
\be
\label{eq:CI}
v_u \, Y_D 
= i \, U^\ast \, \sqrt{m_\nu^{\rm diag}} \, 
R \, \sqrt{M_R^{\rm diag}} 
= i \, \sqrt{v_R}  \, U^\ast \, 
\sqrt{m_\nu^{\rm diag}} \, R \, \sqrt{f_R^{\rm diag}} \, .
\ee
Here $R$ is a complex orthogonal matrix that contains the 
parameters which are integrated out when $m_\nu$ is 
obtained, and which therefore cannot be 
determined from low energy neutrino data without additional 
input. Many analyses of neutrino mixing, 
LFV and/or leptogenesis in the \mbox{type I} 
seesaw framework have been performed using this parameterization 
\cite{LFVIana,PPTS}. 
However, to the best of our knowledge, no 
parameterization of this kind 
has been suggested for the general case of 
\mbox{type I\,+\,II}  seesaw. 
The purpose of the present paper is to generalize the 
Casas-Ibarra 
parameterization to the case of the combined 
\mbox{type I\,+\,II}  seesaw, 
when the mass matrix of light neutrinos 
is given by Eq.~(\ref{eq:II}), 
and to demonstrate the usefulness of the 
proposed parameterization. 

The paper is organized as follows. We summarize the main aspects of lepton 
flavor violation, leptogenesis and neutrino mixing within \mbox{type I\,+\,II}  
seesaw in Section~\ref{sec:II}. Section~\ref{sec:main} contains our central 
results. Here we introduce our parameterization of the Dirac-type Yukawa 
coupling matrix in the case of the combined \mbox{type I\,+\,II}  seesaw.  
We also give simple examples on its usage, two of which are based on the 
approximate tri-bimaximality of neutrino mixing. We conclude in 
Section~\ref{sec:concl}.

\section{\label{sec:II}Formalism}

We will work in the basis in which the mass matrix of 
charged leptons is real and diagonal. The mass 
matrix of light neutrinos is then diagonalized 
according to Eq.~(\ref{eq:mnu}), with $U$ the leptonic 
mixing matrix, for which we will use the parameterization
\begin{equation} 
\label{eq:Upara}
 U =  \left(
 \begin{array}{ccc}
 c_{12} \, c_{13} & s_{12}\, c_{13} & s_{13}\, e^{-i \delta}\\
 -c_{23}\, s_{12}-s_{23}\, s_{13}\, c_{12}\, e^{i \delta} &
 c_{23}\, c_{12}-s_{23}\, s_{13}\, s_{12}\, 
e^{i \delta} & s_{23}\, c_{13}\\
 s_{23}\, s_{12}-\, c_{23}\, s_{13}\, c_{12}\, e^{i \delta} &
 -s_{23}\, c_{12}-c_{23}\, s_{13}\, s_{12}\, 
e^{i \delta} & c_{23}\, c_{13}
 \end{array}
 \right) P \, . 
\end{equation}
Here $c_{ij} = \cos \theta_{ij}$, $s_{ij} = \sin \theta_{ij}$,  
$\delta$ is the Dirac-type CP-violating phase, 
and the Majorana phases 
$\alpha$ and $\beta$ are contained in the matrix 
\be
P = {\rm diag}(1, ~e^{-i \alpha}, ~e^{-i \beta})\,. 
\label{eq:P}
\ee
The analyses of neutrino experiments revealed the 
following best-fit values 
and $3\sigma$ ranges of the oscillation 
parameters \cite{newdata}: 
\begin{eqnarray}
\dms \equiv m_2^2 - m_1^2 &=& 
\left(7.9^{+1.1}_{-0.89}\right) 
\cdot 10^{-5} \eV^2\, ,\nonumber\\
\sss &=& 0.31^{+0.07}_{-0.05} \, ,\nonumber\\
\dma \equiv \left| m_3^2 - m_1^2 \right|&=&  
\left(2.6^{+0.6}_{-0.6}\right) 
\cdot 10^{-3} \eV^2\, ,\\
\sin^2\theta_{23} &=& 0.47^{+0.17}_{-0.15} \, ,\nonumber\\
|U_{e3}|^2  &=&0^{+0.040}_{-0.000} \, .\nonumber
\end{eqnarray}
Depending on the sign of $m_3^2 - m_1^2$, 
the neutrino masses are normally 
or inversely ordered: 
\be \nonumber
\label{eq:masses}
\bav
\text{normal:}  & m_3 > m_2 > m_1 & \text{with} 
& m_2 = \sqrt{m_1^{2}+\dms} ~;
~~~~~~  m_3 = \sqrt{m_1^{2}+\dma} \, ,\\[0.3cm]
\text{inverted:} & m_2 > m_1 > m_3 &  \text{with} & 
m_2 = \sqrt{m_3^{2}+\dms+\dma} ~;~~~~ 
m_1 = \sqrt{m_3^{2} + \dma}\,.
\ea
\ee
The overall scale of neutrino masses is not known, except 
for the upper limit of order 1~eV coming from direct mass search 
experiments and cosmology. 

The type I seesaw mechanism \cite{I} corresponds 
to the situation when  
the light neutrino masses are induced by their coupling 
with heavy Majorana 
neutrinos. Introducing the Dirac mass matrix 
$m_D = v_u \, Y_D$ with 
$v_u = v \, \sin \beta$ being the VEV of $H_u$, 
and the Majorana mass matrix 
$M_R = v_R \, f_R$ for heavy neutrinos, one 
finds for $v_R \gg v_u $ 
the light neutrino mass matrix 
\be 
\label{eq:mnuI}
m_\nu^I = - \frac{v_u^2}{v_R} \,   
\, Y_D \, f_R^{-1} \, Y_D^T \, .
\ee
The masses of light neutrinos can also be 
generated through their coupling 
with an $SU(2)_L$-triplet Higgs, which 
gives the triplet, or type II 
seesaw \cite{II}: 
\be \label{eq:mnuII}
m_\nu^{II} = v_L \, f_L\,.
\ee
In this case the neutrino mass matrix is directly given 
by the triplet 
Yukawa coupling matrix $f_L$, up to an overall 
scale which is just the 
triplet  VEV $v_L$. In left-right 
symmetric models and their GUT extensions 
both seesaw contributions to $m_\nu$ are naturally present, 
leading to 
the \mbox{type I\,+\,II}  seesaw expression of Eq.~(\ref{eq:II}). 
Moreover, in these 
models there is a relation between the VEVs of 
the neutral components of the 
two triplets $v_L$ and $v_R$: 
\be
\label{eq:gamma}
v_L \, v_R = \gamma \, v_u^2\,,
\ee
where $\gamma$ depends on the parameters 
of the Higgs potential. 
Type I + II seesaw can, of course, also be 
realized  without extending the 
gauge group of the standard model. 

In general, the 
matrices $m_\nu$, $f_L$ and $f_R$ are complex 
symmetric, whereas $Y_D$ 
is a general complex matrix (of dimension $3\times 3$ for three  
generations of light and heavy neutrinos). 
In left-right symmetric models and their extensions, in 
addition to the gauge 
symmetry, a discrete left-right symmetry is often assumed, 
which can be 
realized either as $C$-conjugation or as a parity symmetry. 
This leads to 
additional constraints on the entries of the 
seesaw relation~(\ref{eq:II}). 
Namely, in the case of $C$-conjugation symmetry, 
one has $f_L = f_R$, 
$Y_D=Y_D^T$, while for parity 
symmetry $f_L = f_R^\ast$, $Y_D=Y_D^\dag$. In both 
cases the seesaw exhibits a curious duality 
property \cite{AF1} (see also 
\cite{Host,AF2}). In our study, however, we will 
not assume any additional  
constraints on the entries of Eq.~(\ref{eq:II}). 
As the neutrino mass matrix given by this formula 
contains two terms, it 
leads to a number of interesting possibilities 
for explaining the features of 
neutrino mixing \cite{LR}.

Let us now briefly summarize the LFV formulae 
relevant to our discussion. 
In supersymmetric scenarios LFV  is triggered 
by off-diagonal entries 
in the slepton mass matrix $\tilde{m}_L^2$. 
The branching ratios for radiative decays of the charged leptons 
$\ell_i = e, \mu, \tau$ are 
\be \label{eq:BR}
{\rm BR}(\ell_i \ra \ell_j \gamma) 
= {\rm BR}(\ell_i \ra \ell_j \, \nu 
\bar{\nu}) \, \frac{\alpha^3}{G_F^2 \, m_S^8} 
\, \left| \left( \tilde{m}_L^2 \right)_{ij} \right|^2 
\, \tan^2 \beta\, ,
\ee
where $m_S$ is a typical mass scale of SUSY particles. 
The values of the 
branching ratios ${\rm BR}(\ell_i \ra \ell_j \, \nu \bar{\nu})$ 
are ${\rm BR}(\mu \ra e \, \nu \overline{\nu}) \simeq 1$, 
${\rm BR}(\tau \ra \mu \, 
\nu \bar{\nu})\simeq 0.174$ and ${\rm BR}(\tau \ra e \, \nu 
\overline{\nu}) \simeq 0.178$ \cite{PDG}. Current limits 
on the branching ratios for $\ell_i \ra \ell_j \gamma$ are 
BR$(\mu \ra e \gamma) \le 1.2 \cdot 10^{-11}$ \cite{mueg_lim}, 
${\rm BR}(\tau \ra e \gamma) \le 
1.1 \cdot 10^{-7}$ \cite{teg_lim} and 
${\rm BR}(\tau \ra \mu \gamma) \le 6.8 \cdot 10^{-8}$ 
\cite{tmg_lim}. One expects to improve these bounds by 
two to three orders of magnitude for BR$(\mu \ra e \gamma)$ 
\cite{meg_fut} and by one to two orders of magnitude 
for the other branching ratios \cite{BR_fut}. 

To satisfy the requirement that the LFV branching ratios 
${\rm BR}(\ell_i \ra \ell_j \gamma)$ be below their 
experimental upper 
bounds, one typically assumes that $\tilde{m}_L^2$ and all other 
slepton mass and trilinear coupling matrices are 
diagonal at the scale $M_X$. 
Such a situation occurs for instance in mSUGRA scenarios. 
Off-diagonal terms 
get induced at low energy scales radiatively, which 
explains their smallness. In this case a very good 
approximation for the typical SUSY mass appearing in 
Eq.~(\ref{eq:BR}) is 
$m_S^8 = 0.5 \, m_0^2 \, m_{1/2}^2 \, 
(m_0^2 + 0.6 \, m_{1/2}^2)^2$ \cite{PPTS}, where $m_0$ is 
the universal 
scalar mass and $m_{1/2}$ is the universal 
gaugino mass at $M_X$. In a supersymmetric seesaw framework 
the radiative entries giving rise to LFV depend on the same parameters as 
the neutrino masses. If there is only the type I seesaw term 
in $m_\nu$, the well-known result is \cite{ILFV} 
\be \label{eq:LFVI}
\left(\tilde{m}_L^2 \right)_{ij}^I 
= - \frac{(3 m_0^2 + A_0^2)}
{8 \, \pi^2} \, 
\left( Y_D \, L \, Y_D^\dagger \right)_{ij}\,,~\mbox{ where } ~\;
L_{ij} = \delta_{ij} \, \ln \frac{M_X}{M_i} \,.
\ee
In case when only the triplet term $m_\nu^{II}$ 
contributes to $m_\nu$, 
one finds \cite{IILFV}
\be\label{eq:LFVII}
\left(\tilde{m}_L^2 \right)_{ij}^{II} =  - 3 \, 
\frac{(3 m_0^2 + A_0^2)}{8 \, \pi^2} \, 
\left( f_L \, f_L^\dagger \right)_{ij} \, 
\ln \frac{M_X}{M_\Delta}\,.
\ee
Here and in Eq.~(\ref{eq:LFVI}) $A_0$ is the universal trilinear 
coupling. When both terms in the mass matrix 
Eq.~(\ref{eq:II}) are present, 
their contributions to $\left(\tilde{m}_L^2 \right)_{ij}$ sum up:   
\bea 
\label{eq:LFVI+II}
\hspace*{-9.5cm}
\left(\tilde{m}_L^2 \right)_{ij} = 
\left(\tilde{m}_L^2 \right)_{ij}^I + \left(\tilde{m}_L^2 
\right)_{ij}^{II} \vspace*{0.3cm}\\ 
\D\; = - \, 
\frac{(3 m_0^2 + A_0^2)}{8 \, \pi^2} \, 
\left[ 
\left( Y_D \right)_{ik} \,\left(Y_D^\dagger \right)_{kj} \, 
\ln \frac{M_X}{M_k}
+ 3 \, \left( f_L \, f_L^\dagger \right)_{ij} \, 
\ln \frac{M_X}{M_\Delta}
\right] \, .
\eea
As the LFV branching ratios depend on the absolute value 
squared of this 
quantity, there will be a interference term between 
the contributions from the 
triplet term and from the type I seesaw term if both of 
them have off-diagonal entries. 
We will now compare the structures of two expressions: 
\be
\bas 
m_\nu &=& v_L \, f_L  - 
\frac{\D v_u^2}{\D v_R} \, Y_D \, f_R^{-1} \, Y_D^T 
& \mbox{ versus }\, & 
\left(\tilde{m}_L^2 \right)_{ij} &\propto&  
\left(f_L \, f_L^\dagger \right)_{ij} + 
 \left(Y_D \, Y_D^\dagger \right)_{ij}\,,
\ea
\ee
where we have omitted logarithmic corrections 
to $\left(\tilde{m}_L^2 
\right)_{ij}$. There are several possibilities, depending on the 
relative magnitudes of the two contributions to 
$\left(\tilde{m}_L^2 
\right)_{ij}$ and $m_\nu$: 
\begin{itemize}
\item[(i)] in the neutrino mass matrix the type 
I seesaw term $m_\nu^I$ 
dominates, and in the off-diagonal entries of the 
RG-induced slepton mass 
matrix $\left(\tilde{m}_L^2 \right)_{ij}^I$ dominates. 
This situation is the 
one best studied in the literature (\cite{CI,LFVIana,PPTS}, 
for a recent review see \cite{LFV_rev}). We have nothing new to 
add in this case;  

\item[(ii)]in the neutrino mass matrix the triplet term 
$m_\nu^{II}$ dominates, 
and in the off-diagonal entries of the RG-induced slepton 
mass matrix  
$\left(\tilde{m}_L^2 \right)_{ij}^{II}$ dominates. 
This situation has also 
been studied \cite{Rossi,LFVIIana}, though less often than (i);

\item[(iii)]in the neutrino mass matrix the triplet 
term $m_\nu^{II}$ 
dominates, while in the off-diagonal entries of the 
RG-induced slepton mass 
matrix $\left(\tilde{m}_L^2 \right)_{ij}^{I}$ dominates. 
This situation, to our knowledge, has not been studied yet. 
However, there are hardly any 
useful statements to be made, as there is no link between 
neutrino masses and 
LFV, even if $f_L$ and $f_R$ are related by 
$f_L = f_R$ or $f_L = f_R^\ast$; 

\item[(iv)]in the neutrino mass matrix the conventional 
seesaw term 
$m_\nu^{I}$ dominates, whereas in the off-diagonal 
entries of the RG-induced 
slepton mass matrix $\left(\tilde{m}_L^2 \right)_{ij}^{II}$ 
dominates. 
Again, this situation remains to be investigated. 
However, as in case (iii), 
there is hardly any link between neutrino masses 
and LFV, even if $f_L$
and  $f_R$ are related;   

\item[(v)] both terms are of comparable magnitude both in 
$m_\nu$ and in the off-diagonal entries of the 
slepton mass matrix. This case will be of prime interest to us. 

\end{itemize}

In the next section we will propose a Yukawa 
coupling parameterization to 
deal with case (v), which in principle can also 
be applied to cases (iii) and (iv).

Before we turn to the Yukawa coupling parameterization, 
let us summarize the relevant leptogenesis formulae. We 
will assume, as 
it has been done in most studies, that the heavy 
Majorana neutrinos are 
lighter than the Higgs triplets. In that case it 
is sufficient to consider 
only the decay of heavy neutrinos into lepton and 
Higgs doublets (and 
similarly for the SUSY partners), while the decays 
of the triplets into two 
lepton doublets can be neglected. The CP-violating 
decay asymmetries of 
heavy neutrinos $N_i$ contain two contributions. 
The first one is the same 
as in the case of pure type I seesaw \cite{CRV,D}: 
\bea 
\label{eq:epsI}
\left(\varepsilon_i^\alpha \right)_N \D  
\,=\, \D \frac{1}{8 \pi } \, \frac{1}{(Y_D^\dagger \, Y_D)_{ii}}  
\, \sum\limits_{j \neq i} 
{\rm Im} \,\Big[ (Y_D^\dagger)_{i \alpha} \, 
(Y_D)_{\alpha j} \, 
\big(Y_D^\dagger \, Y_D \big)_{i j}\Big] \, f(M_j^2/M_i^2)~~~~~~~ \\ 
\D \,~~~+\, 
 \frac{1}{8 \pi } \, 
\frac{1}{(Y_D^\dagger \, Y_D)_{ii}}  
\, \sum\limits_{j \neq i} 
{\rm Im} \Big[ (Y_D^\dagger)_{i \alpha} \, 
(Y_D)_{\alpha j} \, 
\big(Y_D^\dagger \, Y_D \big)_{j i} \Big]\, \frac{1}{1-M_j^2/M_i^2}\,,
\eea
where    
\be
\D f(x) = 
\sqrt{x} \, \Big[ 
\frac{2}{1 - x} - \ln \Big( \frac{1+x}{x} \Big) 
 \Big] \,.
\ee
We have indicated here that flavor 
effects \cite{flavor_flav} might play a 
role, i.e.,~$\varepsilon_i^\alpha$ describes 
the decay of the heavy neutrino 
of mass $M_i$ into leptons of flavor 
$\alpha = e, \mu, \tau$. We will focus 
here on the case when the lowest-mass heavy neutrino 
is much lighter than 
the other two, i.e.~$M_1 \ll M_{2,3}$; the lepton 
asymmetry is then dominated 
by the decay of this lightest neutrino. In this case 
$f(M_j^2/M_1^2) \simeq - 3 \, M_1/M_j$, and in addition the 
second term in Eq.~(\ref{eq:epsI}) is 
strongly suppressed, therefore we will neglect it 
in what follows.  
 
The second type of asymmetry is encountered when a 
Higgs triplet is exchanged 
in the loop diagrams \cite{HamSen}:
\be 
\label{eq:epsII}
\left(\varepsilon_i^\alpha \right)_\Delta \D 
= \frac{3 }{8 \pi} \, \frac{M_i \, v_L}{v_u^2} 
\, \frac{1}{(Y_D^\dagger \, Y_D)_{ii}} \, 
{\rm Im} \Big[ 
\big( f_L^\ast \, Y_D \big)_{\alpha i} \, (Y_D^T)_{i \alpha} 
\Big] \, g(M_\Delta^2 /M_i^2)\,, 
\vspace*{-2mm}
\ee
where    
\be
\D g(x) = x \, \ln \left( \frac{1+x}{x} \right)\,.
\ee 
In the limit $M_\Delta \gg M_{1,2,3}$, which we 
will assume, one has  
$g(x) \simeq 1 - \frac{1}{2 \, x} 
= 1 - \frac 12 \, (M_i/M_\Delta)^2$. 
The total asymmetries $(\varepsilon_i)_N$ 
and $(\varepsilon_i)_\Delta$ 
are obtained by summing $(\varepsilon_i^\alpha)_N$ and 
$(\varepsilon_i^\alpha)_\Delta$ over the flavor index 
$\alpha$. 

The baryon asymmetry of the universe 
($\eta_B = n_B/n_\gamma = 6.1 
\cdot 10^{-10}$) is finally found as 
\be 
\label{eq:tab}
\eta_B \simeq -0.96 \cdot 10^{-2} 
\sum\limits_\alpha \varepsilon_1^\alpha \, \kappa^\alpha\,,
\ee 
where the washout factors $\kappa^\alpha$ are 
obtained by solving the 
relevant Boltzmann equations. The approximate 
expression we use is \cite{PB}
$$
 \kappa^\alpha \simeq \frac{2}{K^\alpha \, z_B(K^\alpha)} \big\{
 1 - \exp{\big[-K^\alpha \, z_B(K^\alpha)/2\big] }
 \big\}\,,
$$
where $K^\alpha = \sum K_i^\alpha$ with 
$K_i^\alpha = |(Y_D)_{\alpha 
i}|^2\, K_i /(Y_D^\dagger \, Y_D)_{ii}$ and 
\be
z_B(K^\alpha) = 2 + 4\,(K^\alpha)^{0.13}\,
\exp \left(-2.5/K^\alpha \right)\,.
\ee
The parameter $K_i$ in the expression for $K_i^\alpha$ 
is defined as  
$K_i = \left. \Gamma_i/H(T)\right|_{T = M_i}$, with 
the tree-level decay width 
of the $i$th heavy neutrino 
$\Gamma_i = (Y_D^\dagger \, Y_D)_{ii} \, 
M_i/(8 \, \pi)$ and the Hubble 
parameter $H(T) = 1.66 \, \sqrt{g^\ast} \, T^2 
/M_{\rm Pl}$. The out-of-equilibrium decay condition for $N_i$ is 
essentially $K_i < 1$.

\section{\label{sec:main}Dirac-type Yukawa coupling 
parameterization for 
\mbox{\mbox{type I\,+\,II} } seesaw and its applications}

When the triplet term $m_\nu^{II}$ is present in 
the seesaw relation, 
the procedure that led to the Casas-Ibarra parameterization 
(\ref{eq:CI}) of the matrix $Y_D$ cannot be directly applied. 
However, as we 
shall show, a simple transformation of Eq.~(\ref{eq:II}) 
makes it possible 
to generalize the parameterization (\ref{eq:CI}) to 
the case of \mbox{type I\,+\,II}  
seesaw. 

First, we move the type II contribution to the left hand side of 
Eq.~(\ref{eq:II}), which gives 
\be
\label{eq:IIa}
m_\nu - v_L \, f_L = - 
\frac{v_u^2}{v_R} \, Y_D \, f_R^{-1} \, Y_D^T\,. 
\ee
It is convenient to introduce the notation 
\be
\label{eq:Xnu}
X_\nu \,\equiv\, m_\nu - v_L \, f_L\,,\quad 
\mbox{diagonalized as}\, \quad
X_\nu \,=\, V_\nu^\ast \, X_\nu^{\rm diag} V_\nu^\dagger
\ee
with a unitary matrix $V_\nu$. 
Multiplying both sides of 
Eq.~(\ref{eq:IIa}) by $X_\nu^{-1/2}$, we find 
\be
\label{eq:IIb}
\mathbbm{1}\,=\, - \frac{v_u^2}{v_R} \, 
\big(X_\nu^{-1/2}\, Y_D \, f_R^{-1/2}
\big) \, \big(f_R^{-1/2} \, Y_D^T\, X_{\nu}^{-1/2}\big)\,. 
\ee 
Next, we note that although a square root of a symmetric 
matrix is not 
always automatically symmetric, it can always be chosen 
to be symmetric. 
We will make such a choice for the matrices $X_\nu^{-1/2}$ and 
$f_R^{-1/2}$, i.e.~we assume them to be symmetric.%
\footnote{This can, e.g.,~be achieved by diagonalizing 
$X_\nu$ and $f_R$ by complex orthogonal transformations 
and then taking square roots. }
One can then rewrite Eq.~(\ref{eq:IIb}) as 
\be
\label{eq:IIc}
\mathbbm{1}\,=\, R R^T \qquad \mbox{with} \qquad R\,=\,
\pm i \frac{v_u}{\sqrt{v_R}} \big(X_\nu^{-1/2}\, 
Y_D \, f_R^{-1/2}\big)\,.
\ee
Eq.~(\ref{eq:IIc}) means that the \mbox{type I\,+\,II} 
seesaw relation requires 
$R$ to be an (in general complex) orthogonal matrix, but 
otherwise does 
not constrain it. Thus, for the Dirac-type Yukawa 
coupling $Y_D$ we have 
\be 
\label{eq:paraIIgen}
v_u \, Y_D \,=\, \pm i \, \sqrt{v_R}\; X_\nu^{1/2} \, 
R \; f_R^{1/2}\,,
\ee
where $R$ is an arbitrary complex orthogonal matrix. 
It can be parameterized as 
\be
\label{eq:R}
R \,=\,R_{12} \, R_{13} \, R_{23}\,,
\ee 
where $R_{ij}$ is the matrix of rotation by a complex angle 
$\omega_{ij} = \rho_{ij} + i \, \sigma_{ij}$ in the $ij$-plane. 
The parameterization of the Yukawa coupling matrix $Y_D$ in 
Eq.~(\ref{eq:paraIIgen}) is the most general one satisfying the combined type 
I + II seesaw formula.

As was pointed out above, when the underlying 
theory possesses a discrete 
left-right symmetry, \mbox{type I\,+\,II}  seesaw 
exhibits a duality property \cite{AF1}. 
In that case the seesaw relation~(\ref{eq:II}) is 
invariant with respect to the duality transformation 
$f_R\to \hat{f}_R \equiv m_\nu/v_L - f_L$. 
It is interesting to note that in terms of 
$f_R$ and its dual $\hat{f}_R$ 
Eq.~(\ref{eq:paraIIgen}) can be rewritten as 
\be 
\label{eq:paraIIdual}
v_u \, Y_D \,=\, \pm i \, \sqrt{v_L v_R}\; \hat{f}_R^{1/2} \, 
R \; f_R^{1/2}\,. 
\ee

For practical applications, it proves to be convenient to use a 
slightly modified version of Eq.~(\ref{eq:paraIIgen}). 
First, we note that 
for discussions of both LFV and leptogenesis one has to go to the 
basis where the mass matrix of heavy Majorana neutrinos 
$M_R$ is diagonal 
and real. As $M_R=f_R\, v_R$, this also diagonalizes 
the matrix $f_R$. The 
corresponding transformation is 
\be
\label{eq:trans}
V_R^T \, f_R \, V_R = f_R^{\rm diag}\,,
\ee
with a unitary matrix $V_R$. Note that 
$f_R^{\rm diag} = {\rm diag}(M_1,\,
M_2,\, M_3)/v_R$. The transformation (\ref{eq:trans}) amounts to 
replacing the Yukawa coupling matrix 
$Y_D$ in the seesaw relation 
according to $Y_D \rightarrow Y_D \, V_R$, i.e.~it fixes 
its right-handed basis. In what follows we will be assuming that the matrix 
$M_R$ has been diagonalized and consider $Y_D$ in this basis, except in 
example 3 below, where this diagonalization will be carried out 
explicitly.  
Next, it is convenient to express $X_\nu$ through its eigenvalues.  
To this end, using Eq.~(\ref{eq:Xnu}) we rewrite $X_\nu$ on 
the left hand side of Eq.~(\ref{eq:IIa}) as 
\be
X_\nu \,=\, V_\nu^\ast\,X_\nu^{\rm diag}V_\nu^\dagger 
\,=\,\big[V_\nu^\ast\,(X_\nu^{\rm diag})^{1/2}\big]
\big[V_\nu^\ast\,(X_\nu^{\rm diag})^{1/2}\big]^T\,. 
\ee
Multiplying then Eq.~(\ref{eq:IIa}) by 
$[V_\nu^\ast\,(X_\nu^{\rm diag})^{1/2}
]^{-1}$ on the left and by 
$\{[V_\nu^\ast\,(X_\nu^{\rm diag})^{1/2}]^{T}
\}^{-1}$ on the right and following 
the same steps as above, one readily finds 
\be 
\label{eq:paraII}
v_u \, Y_D = \pm 
i \, \sqrt{v_R} \, V_\nu^\ast \, \sqrt{X_\nu^{\rm diag}} \; 
R \; \sqrt{f_R^{\rm diag}}\,. 
\ee
This parameterization is the main point of the present paper 
and we will be using it in the subsequent 
discussion. Note that the matrix $R$ here is in general 
not the same as the matrix $R$ in Eq.~(\ref{eq:paraIIgen}). This is of 
no concern to us, as both are arbitrary complex symmetric matrices.

In the remainder of this section we will give simple examples demonstrating 
the usefulness of the parameterization~(\ref{eq:paraII}). 
In the first two examples we consider tri-bimaximal 
neutrino mixing \cite{tri}, which describes very 
well the current status of 
global fits to the low energy neutrino data. The 
neutrino mass matrix giving 
rise to tri-bimaximal mixing can be written as  
\be \label{eq:mnutbm}
m_\nu = \frac{m_1}{6} \, 
\left( 
\bad 
4 & -2 & -2 \\
\cdot & 1 & 1 \\
\cdot & \cdot & 1 
\ea 
\right) 
+ \frac{m_2 \, e^{2i \alpha}}{3} \, 
\left( 
\bad 
1 & 1 & 1 \\
\cdot & 1 & 1 \\
\cdot & \cdot & 1 
\ea 
\right) + 
\frac{m_3 \, e^{2i \beta}}{2} \, 
\left( 
\bad 
0 & 0 & 0 \\
\cdot & 1 & -1 \\
\cdot & \cdot & 1 
\ea 
\right)\,.
\ee
If neutrinos enjoy the normal mass hierarchy, 
one can neglect $m_1$, so that 
the first term in Eq.~(\ref{eq:mnutbm}) vanishes, 
and in addition one has 
$m_2 = \sqrt{\dms}$ and $m_3 = \sqrt{\dma}$. An 
appealing possibility in this 
case is that the two remaining individual 
matrices in Eq.~(\ref{eq:mnutbm}) 
correspond to $m_\nu^I$ and $m_\nu^{II}$, 
respectively \cite{LR}. The moderate 
ratio of the two terms in $m_\nu$ is 
therefore $\frac 32 \, \sqrt{\dma/\dms} 
\simeq 8.4$.  
We will investigate this 
possibility and apply our parameterization of $Y_D$ to this case 
in the following two examples.  
The third example will be based on a perturbation of bimaximal leptonic  
mixing \cite{bima} in the type I + II seesaw framework.

\subsection{\label{sec:first}First example}
Suppose first that the triplet term $m_\nu^{II}$ is the term 
proportional to $m_3$ in Eq.~(\ref{eq:mnutbm}), i.e.   
\be
\label{eq:fL}
f_L = 
\left( 
\bad 
0 & 0 & 0 \\
\cdot & 1 & -1 \\
\cdot & \cdot & 1 
\ea 
\right)\, e^{2i \beta}~~\mbox{ and }~v_L =  \sqrt{\dma}/2\, .
\ee
The second, flavor democratic term proportional to $m_2$, is then 
provided by the conventional type I seesaw. Due to the 
seesaw relation 
(\ref{eq:IIa}) it determines $X_\nu$: 
$$
X_\nu = - \frac{v_u^2}{v_R} \, Y_D \, f_R^{-1} \, Y_D^T 
= \frac{m_2\, e^{2i \alpha}}{3} 
\left( 
\bad 
1 & 1 & 1 \\
\cdot & 1 & 1 \\
\cdot & \cdot & 1 
\ea
\right)\,.
$$
Consequently, one can write 
$X_\nu^{\rm diag} = {\rm diag}(0,0,m_2)$ and 
\be
V_\nu = 
\left( 
\bad 
1/\sqrt{2} & 1/\sqrt{6} & 1/\sqrt{3} \\[0.2cm] 
-1/\sqrt{2} & 1/\sqrt{6} & 1/\sqrt{3} \\[0.2cm]
0 & -2/\sqrt{6} & 1/\sqrt{3}
\ea
\right) \, {\rm diag}(1,1,e^{-i \alpha})\,. 
\ee
The scales involved are $v_L \simeq 0.025$ eV, 
$v_R = 3 \, v_u^2 /\sqrt{\dms} 
\simeq 1.0 \cdot 10^{16}$ GeV (assuming $v_u = 174$ GeV, 
which is an excellent approximation 
as long as $\tan\beta \gs 5$), and $\gamma \simeq 8.4$. 
Note that we have 
rather arbitrarily decomposed the second and third terms in 
Eq.~(\ref{eq:mnutbm}) into the VEVs and Yukawa couplings or their 
combinations.  

We have now all ingredients to express 
$Y_D$ through Eq.~(\ref{eq:paraII}), 
and the result is 
\be \label{eq:yd1}
\hspace{-.14cm}Y_D = -i e^{i \alpha} \, \frac{\sqrt{m_2}}
{\sqrt{3} \, v_u} \, 
\left( 
\bad 
\sqrt{M_1} \, \sin \omega_{13} & 
\sqrt{M_2} \, \cos \omega_{13} \, \sin \omega_{23} & 
-\sqrt{M_3} \, \cos \omega_{13} \, \cos \omega_{23} \\[0.2cm]
\sqrt{M_1} \, \sin \omega_{13} & 
\sqrt{M_2} \, \cos \omega_{13} \, \sin \omega_{23} & 
-\sqrt{M_3} \, \cos \omega_{13} \, \cos \omega_{23} \\[0.2cm]
\sqrt{M_1} \, \sin \omega_{13} & 
\sqrt{M_2} \, \cos \omega_{13} \, \sin \omega_{23} & 
-\sqrt{M_3} \, \cos \omega_{13} \, \cos \omega_{23} 
\ea 
\right)\, .
\ee
Interestingly, the complex angle $\omega_{12}$ drops out of this 
expression. 

Let us now discuss LFV in the considered example. 
Eq.~(\ref{eq:fL}) yields  
\be
f_L \, f_L^\dagger = 2 \, 
\left( 
\bad 
0 & 0 & 0\\[0.2cm]
0 & 1 & -1\\[0.2cm]
0 & -1 & 1 
\ea
\right)\,,
\ee
from which it follows that only the decay $\tau \ra \mu \gamma$ 
is influenced by the triplet term. The decays $\mu \ra e \gamma$ 
and $\tau \ra e \gamma$ depend only on $Y_D \, Y_D^\dag$, which 
has a democratic structure with all terms equal to each other. 
Consequently, $\mu \ra e \gamma$ and $\tau \ra e \gamma$ 
decays are governed by the same quantity: 
\bea 
\D \label{eq:meg1}
 \big| \big(Y_D \, Y_D^\dagger\big)_{21} + 3  
\big(f_L \, f_L^\dagger \big)_{21}  \big|^2 
= \big|\big(Y_D \, Y_D^\dagger\big)_{31} + 3  
\big(f_L \, f_L^\dagger \big)_{31}  \big|^2 
\qquad\qquad\qquad~ \\[0.2cm] \D 
= \frac{m_2^2}{9 \, v_u^4} \big( 
M_1 \, |\sin \omega_{13}|^2 + 
|\cos \omega_{13}|^2 \, (M_2 \, |\sin \omega_{23}|^2 
+ M_3 \, |\cos \omega_{23}|^2 ) 
\big)^2\,.
\eea
This equality implies that BR$(\tau \ra e  \gamma) = 0.178 \, 
{\rm BR}(\mu \ra e \gamma)$. With the current limit of $1.2 \cdot 10^{-11}$ 
on BR$(\mu \ra e \gamma)$, and an expected improvement of two 
orders of magnitude on the 
limit of ${\rm BR}(\tau \ra e \gamma) \le 1.1 
\cdot 10^{-7}$, it follows that 
in this scenario $\tau \ra e 
\gamma$ will not be observed in a foreseeable future. The branching ratio 
of the decay $\tau \ra \mu \gamma$ depends on
\be 
\label{eq:tmg1}
\left(Y_D \, Y_D^\dagger\right)_{32} + 3  
\left(f_L \, f_L^\dagger \right)_{32}  
= \left(Y_D \, Y_D^\dagger\right)_{21} - 6 \,.
\ee
We have omitted here the logarithmic dependence on the masses of the triplet 
and of the heavy Majorana neutrinos. In the plots to be shown in the following 
we use the full expressions, however. 
As follows from Eqs.~(\ref{eq:meg1}) and (\ref{eq:tmg1}), 
the matrix $Y_D\,Y_D^\dagger + 3 \, f_L \, f_L^\dagger$ 
depends in general on 
two complex angles, $\omega_{23}$ and $\omega_{13}$. 
If degenerate heavy 
Majorana masses are assumed, $M_1 = M_2 = M_3$, then 
the real part of $\omega_{23}$ drops out of this matrix.

Turning to leptogenesis, the first thing to note is that all 
$(\varepsilon_i^\alpha)_\Delta$ vanish, which is a consequence 
of the fact that the matrix $f_L^\ast \, Y_D$ vanishes 
identically. The decay asymmetry 
is therefore the same as for pure type I seesaw. 
The individual flavored 
asymmetries $(\varepsilon_1^\alpha)_N$ are all 
identical and equal to one third of the total asymmetry. 
For hierarchical heavy neutrinos we find 
\bea 
\D \label{eq:eps1}
(\varepsilon_1^e)_N = (\varepsilon_1^\mu)_N =
 (\varepsilon_1^\tau)_N = 
\frac 13 \, \varepsilon_1^N \simeq 
 \frac{1}{16 \pi} \, \frac{m_2 \, M_1}{v_u^2} \, 
\frac{\sin 2 \rho_{13} ~ \sinh 2 \sigma_{13}}
{|\sin  \omega_{13}|^2} \vspace*{0.3cm}
\\[0.2cm] \D 
\simeq 6 \cdot 10^{-9}  \left(\frac{M_1}{10^9~\rm GeV} \right) 
\frac{\sin 2 \rho_{13} ~ \sinh 2 \sigma_{13}}
{|\sin  \omega_{13}|^2}\,.
\eea
Hence, only the complex angle $\omega_{13}$ plays a role here. Terms 
containing $\omega_{23}$ appear in the decay asymmetry multiplied by  
$f(M_2^2/M_1^2) \, M_2 - f(M_3^2/M_1^2)\, M_3$, which vanishes in the limit 
of hierarchical heavy neutrinos. If $\omega_{13}$ is zero, then $N_1$ 
decouples (see Eq.~(\ref{eq:yd1})), and $N_2$ will be responsible for 
leptogenesis. The low energy (Majorana) phases $\alpha$ and $\beta$ do not 
contribute to either $\varepsilon_1$ or to $\varepsilon_2$, i.e.~play no role 
in leptogenesis.

\begin{figure}[t]
\begin{center}
\epsfig{file=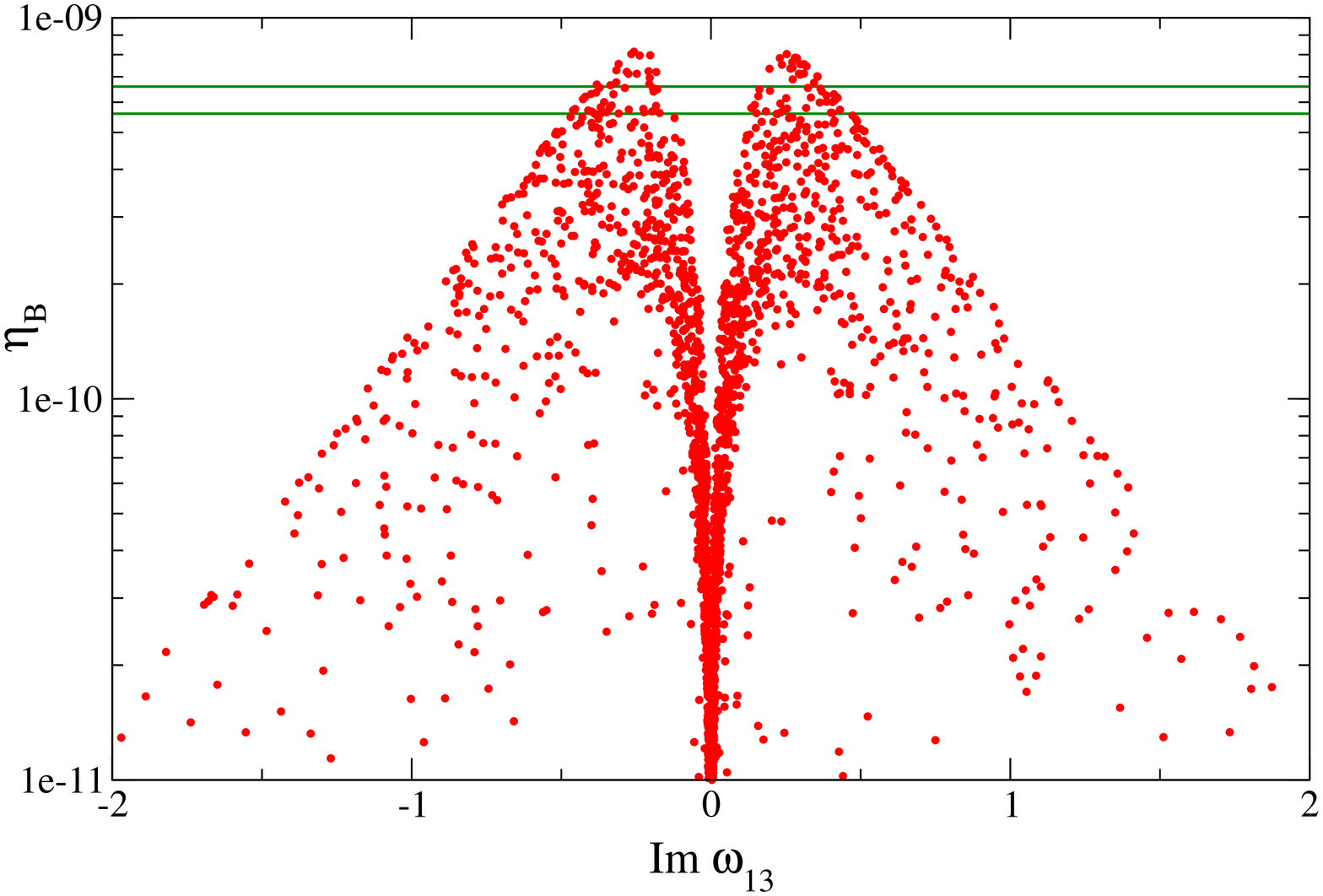,width=12cm,height=8cm}
\caption{\label{fig:ex1}Scatter plot for the baryon 
asymmetry $\eta_B$ against the imaginary part of 
$\omega_{13}$ for the first example of 
Section~\ref{sec:first}. The observed value of 
$\eta_B$ corresponds to the region between the horizontal lines. }
\end{center}
\end{figure}

Choosing $M_X = 2 \cdot 10^{16}$ GeV, 
$M_\Delta = 5 \cdot 10^{15}$ 
GeV and the masses of heavy Majorana neutrinos 
$M_1 = 10^{10}$ GeV, 
$M_2 = 10^{12}$ GeV, and $M_3 = 10^{15}$ GeV, we show 
in Fig.~\ref{fig:ex1} 
the baryon asymmetry against the imaginary part of 
$\omega_{13}$. All free parameters were varied,
the baryon asymmetry was required to be positive and 
the branching ratios of $\mu \ra e \gamma$ 
(which in the considered example coincides with 
${\rm BR}(\tau \ra e \gamma)/0.178$) and of $\tau \ra \mu \gamma$ 
were required to lie below their 
current upper limits. 
The supersymmetric parameters we have used correspond 
to the SPS benchmark point 2 of Ref.~\cite{SPS} and are $\tan \beta = 10$, 
$m_0 = 1450$ GeV, $m_{1/2} = 300$ GeV and $A_0 = 0$. The apparent symmetry 
of Fig.~\ref{fig:ex1} around the value Im$(\omega_{13}) = \sigma_{13} = 0$ 
can be explained by the dependence of the decay asymmetry~(\ref{eq:eps1}) 
on $\omega_{13}$.  
For all other parameters fixed, changing the sign of 
$\sigma_{13}$ would also change the sign of the decay asymmetry. 
To regain the correct sign of the baryon asymmetry one would then also 
have to flip the sign of $\sin 2\rho_{13}$ (recall that $\rho_{13}$ is 
varied as a free parameter in this scatter plot), leading to the apparent 
symmetry of the figure.

\begin{figure}[t]
\begin{center}
\epsfig{file=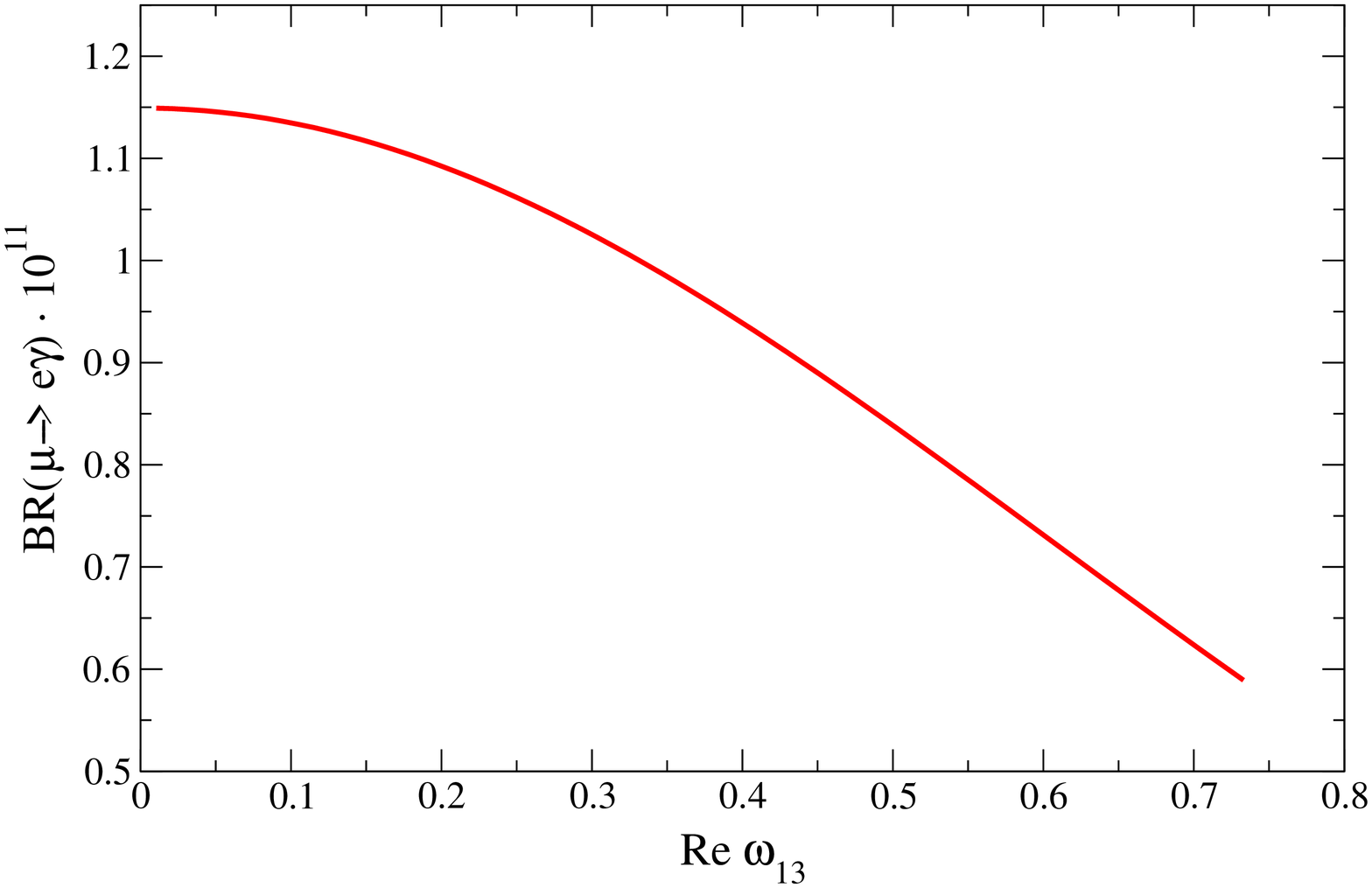,width=12cm,height=8cm}
\caption{\label{fig:ex2}The branching ratio of 
$\mu \ra e \gamma$ decay 
against the real part of $\omega_{13}$ for a particular 
point in the parameter space of the first example of 
Section~\ref{sec:first} (see the 
text for details).}
\end{center}
\end{figure}

\begin{figure}[t]
\begin{center}
\epsfig{file=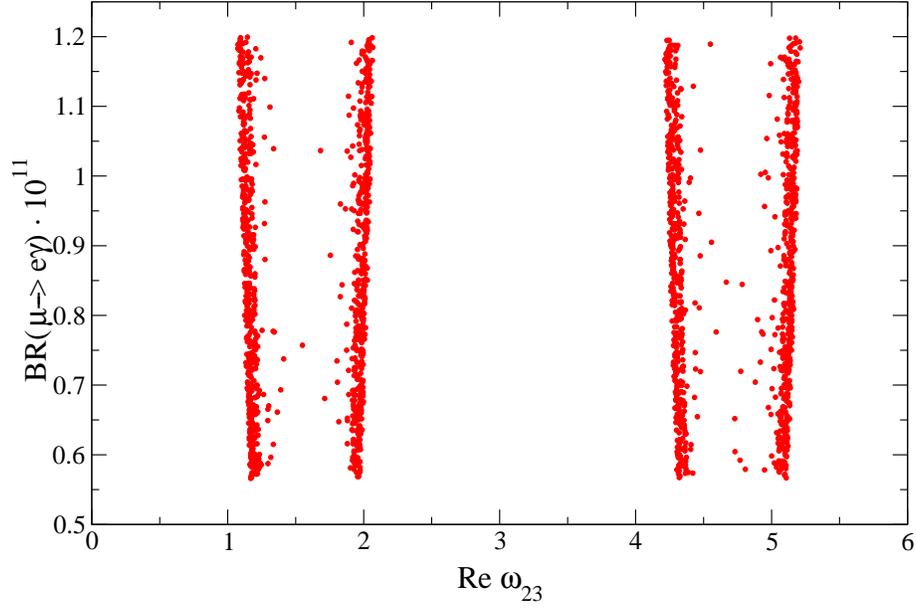,width=12cm,height=8cm}
\caption{\label{fig:ex3}First example from~Section \ref{sec:first}: 
scatter plot for the branching ratio of $\mu \ra e \gamma$ 
decay against the real part of $\omega_{23}$ when 
the baryon asymmetry $\eta_B$ is within its 
experimental range.}
\end{center}
\end{figure}

The branching ratio ${\rm BR}(\tau \ra \mu \gamma)$ is basically 
independent of the parameters of $Y_D$, because the constant term in 
Eq.~(\ref{eq:tmg1}) turns out to be much larger than the $Y_D$-dependent one. 
The ratio ${\rm BR}(\mu \ra e \gamma)$/${\rm BR}(\tau \ra \mu \gamma)$ 
is of order $10^{-4}$, implying that $\tau \ra \mu \gamma$ is observable 
as long as ${\rm BR}(\mu \ra e \gamma)$ is close to its current limit. 
Fixing in addition $\rho_{23} = 1.7$, $\sigma_{23} = -0.3$ and $\sigma_{13} = 
-0.7$, we show in Fig.~\ref{fig:ex2} the branching ratio of $\mu \ra e \gamma$ 
against the remaining free parameter $\rho_{13} = {\rm Re}(\omega_{13})$. 
For this particular point BR$(\tau \ra \mu \gamma)\simeq 5.05 \cdot 10^{-8}$, 
which is very close to its current upper limit. Fig.~\ref{fig:ex3} shows a 
scatter plot for the branching ratio of $\mu \ra e \gamma$ against the real 
part of $\omega_{23}$ when the baryon asymmetry is within its allowed range. 
The symmetry around the value Re$(\omega_{23}) = \rho_{23} = \pi$ of this plot 
can be understood by noting that the term proportional to $M_3 \, |\cos 
\omega_{23}|^2$ is the leading one in Eq.~(\ref{eq:meg1}). This term 
depends on $\rho_{23}$ through $\cos 2 \rho_{23}$.

\subsection{\label{sec:sec}Second example}
Let us now consider the situation in which the triplet 
term is flavor 
democratic, i.e., 
$$
f_L = 
\left( 
\bad 
1 & 1 & 1 \\
\cdot & 1 & 1 \\
\cdot & \cdot & 1 
\ea 
\right)\, e^{2i \alpha}~~\mbox{ and }~v_L =  \sqrt{\dms}/3\,.
$$
The remaining term in Eq.~(\ref{eq:mnutbm}) is then 
$$
X_\nu = - \frac{v_u^2}{v_R } \, Y_D \, f_R^{-1} \, Y_D^T = 
\frac{m_3 \,e^{2 i \beta}}{2}
\left( 
\bad
0 & 0 & 0\\
\cdot & 1 & -1 \\
\cdot & \cdot & 1
\ea
\right)\,.
$$
The involved scales are $v_L \simeq 0.003$ eV, 
$v_R = 2 \, v_u^2 /\sqrt{\dma} 
\simeq 1.2 \cdot 10^{15}$ GeV, and $\gamma \simeq 0.12$.  
Here we have 
taken $v_u^2/v_R = m_3/2=\sqrt{\dma}/2$. 
The matrix $X_\nu$ is diagonalized by  
\be
V_\nu = 
\left( 
\bad 
1 & 0 & 0 \\[0.2cm]
0 & 1/\sqrt{2} & -1/\sqrt{2} \\[0.2cm]
0 & 1/\sqrt{2} & 1/\sqrt{2}
\ea
\right) \, {\rm diag}(1,1,e^{-i \beta})\, . 
\ee
with $X_\nu^{\rm diag} = {\rm diag}(0,0,m_3)$. 
From Eq.~(\ref{eq:paraII}) 
we then obtain 
\be
Y_D = -i \, e^{i \beta} \frac{\sqrt{m_3}}{\sqrt{2} \, v_u} 
\left( 
\bad 
0 & 0 & 0 \\[0.2cm]
\sqrt{M_1} \, \sin \omega_{13} 
& \sqrt{M_2} \, \cos \omega_{13} \, \sin \omega_{23} 
& - \sqrt{M_3} \, \cos \omega_{13} \, \cos \omega_{23} \\[0.2cm]
-\sqrt{M_1} \, \sin \omega_{13} 
& - \sqrt{M_2} \, \cos \omega_{13} \, \sin \omega_{23} 
& \sqrt{M_3} \, \cos \omega_{13} \, \cos \omega_{23} \\[0.2cm]
\ea
\right)\,.
\ee
Note that, as in the previous example, $Y_D$ does not depend on 
$\omega_{12}$. 
The matrix $Y_D Y_D^\dag$ has zero first row and 
first column, therefore $\mu \ra e \gamma$ and $\tau \ra e 
\gamma$ decays are governed by the triplet contribution, 
and depend  
on the the same quantity, namely $(f_L \, f_L^\dagger)_{21} = 
(f_L \, f_L^\dagger)_{31} = 3$ 
(note that $f_L \, f_L^\dagger = 3 \, f_L 
\, e^{-2 i \alpha}$, i.e.~is flavor democratic and has 
no dependence on any of the free parameters). Consequently, the decay $\tau 
\ra e \gamma$ will not be observed in a near future. 
The fact that $\mu \ra e \gamma$ and 
$\tau \ra e \gamma$ decays depend on the same quantity in both 
our examples is a consequence of the $\mu$--$\tau$ symmetry of the 
involved mass matrices. Finally, 
\bea
\left(Y_D \, Y_D^\dagger\right)_{32} + 3  
\left(f_L \, f_L^\dagger \right)_{32} \vspace*{0.1cm} \\[0.2cm] \D 
= \frac{1}{2 \, v_u^2} \, \left| 
18 \, v_u^2 - m_3 \, 
(
M_1 \, |\sin \omega_{13}|^2 + 
|\cos \omega_{13}|^2 (M_2 \, |\sin \omega_{23}|^2 
+ M_3 \, |\cos \omega_{23}|^2 ) 
)
\right|\, .
\eea
Because the 12- and 13-entries of $f_L \, f_L^\dagger$ are 
independent of any free parameters, it is 
not possible to suppress them, and in general the branching ratios of LFV 
decays are too large unless the SUSY masses are around or above 10 TeV. 

Let us now turn to leptogenesis. As in the previous example, all 
$(\varepsilon_i^\alpha)_\Delta$ vanish because the 
matrix $f_L^\ast \, Y_D$ vanishes identically. The decay asymmetry 
$(\varepsilon_1^e)_N$ is also zero, whereas $(\varepsilon_1^\mu)_N$ and 
$(\varepsilon_1^\tau)_N$ are identical, and equal to $\frac 12 \, 
(\varepsilon_1)_N$. In the limit of the hierarchical heavy neutrino masses we 
find 
\bea \D 
(\epsilon_1^\mu)_N = (\varepsilon_1^\tau)_N = \frac 12 \, 
(\varepsilon_1)_N = 
\frac{3}{32 \pi} \, \frac{m_3 \, M_1}{v_u^2} \, 
\frac{\sin 2 \rho_{13} ~ \sinh 2 \sigma_{13}}
{|\sin  \omega_{13}|^2} \vspace*{0.2cm} \\[0.2cm] \D 
\simeq 5 \cdot 10^{-8}  \left(\frac{M_1}{10^9~\rm GeV} \right) 
\frac{\sin 2 \rho_{13} ~ \sinh 2 \sigma_{13}}
{|\sin  \omega_{13}|^2}\, .
\eea
The dependence of these asymmetries on the complex angle $\omega_{13}$ 
is identical to that in the first example considered above. Note that here 
the decay asymmetry is proportional to the mass of the heaviest of 
light neutrinos $m_3$, whereas it was proportional to $m_2$ in the 
first example.

\subsection{\label{sec:third}Third example}

Our final example is based on the following 
observation~\cite{WRII,LR}: 
if the triplet term corresponds to bimaximal mixing \cite{bima} ($U_{e3} = 0$ 
and $\theta_{12} = \theta_{23} = \pi/4$), then a small contribution from the 
conventional type I seesaw term may shift $\theta_{12}$ sufficiently away from 
the maximal mixing value to make it agree with data. 
Non-zero $\theta_{13}$ and non-maximal mixing in the 2-3 sector are also 
generated. It was assumed in \cite{WRII,LR} that $m_D$ is hierarchical, 
symmetric and coincides with the up-type quark mass matrix.
The triplet term $v_L \, f_L$ alone would generate bimaximal neutrino mixing 
and a normal mass ordering with a non-vanishing smallest neutrino mass. 
A discrete left-right symmetry is also assumed, such that $f_L = f_R$. 
It is easy to see that in this case the type I seesaw term contributes 
to $m_\nu$ mainly a $33$ entry $v_L \, \eta$, which is suppressed with 
respect to the leading (order $v_L$) term of $m_\nu^{II}$ \cite{JPR0}.  
The other elements of $m_\nu^I$ are much smaller than 
$v_L \, \eta$, and we will neglect them. It should be noted that many other 
Dirac mass matrices can also give the desired form $m_\nu^{I} \propto 
{\rm diag}(0,0,1)$, and our parameterization allows to study them all.  

The triplet contribution is 
$$
f_L = f_R = 
\left( 
\bad 
\epsilon & B \, \epsilon & B \, \epsilon \\
\cdot & \frac12 \, (\epsilon + e^{i \phi}) & \frac12 \, 
(\epsilon - e^{i \phi}) \\
\cdot & \cdot & \frac12 \, (\epsilon + e^{i \phi}) \\
\ea
\right)\quad\mbox{and}\quad
 v_L = \frac{\sqrt{\dma}}{2}\, .
$$
For simplicity we assume the order one parameter $B$ 
and $\epsilon \ll 1$ to 
be real. The product $v_L \, \epsilon$ is of the order of $\sqrt{\dms}$.  
 
The type I contribution we require is 
$$
X_\nu = -\frac{v_u^2}{v_R} \, Y_D \, f_{R}^{-1} \, Y_D^T 
\equiv v_L  \, 
\left( 
\bad
0 & 0 & 0 \\
\cdot & 0 & 0 \\
\cdot & \cdot & \eta
\ea
\right)\,.
$$
The involved scales are $v_L = 0.025$ eV 
and $v_R = 2 \, v_u^2/\sqrt{\dma} \simeq 
1.2 \cdot 10^{15}$ GeV. As a consequence of 
non-zero $\eta$, the 
zeroth-order values $U_{e3} = 0$ and 
$\theta_{12} = \theta_{23} = \pi/4$ 
are modified to $|U_{e3}| \simeq B \, \epsilon \, \eta/\sqrt{2}$, 
$\tan^2 \theta_{23} \simeq 1 - 2 \, \eta$ and $\tan 2 \theta_{12} 
\simeq 4 \sqrt{2} \, B \, \epsilon/\eta$, where 
for simplicity also $\eta$ is assumed to be real. 
The value $\sin^2 \theta_{12} = 
\frac 13$ is achieved for $B \, \epsilon = \eta/2$. The ratio 
of the neutrino mass squared differences $\dms/\dma$ 
is approximately $\frac 34 \, \eta \, (4 \, \epsilon + \eta)$. 
A choice of parameters which leads to neutrino properties that agree with 
the data, and which we will use in what follows, is $B = 1.1$, 
$\eta = 0.1194$ and $\epsilon = 0.0542$. The low 
energy phase $\phi$ is the Dirac-type CP violation phase 
which can influence neutrino oscillations.

Since $X_\nu$ is diagonal, we have 
$V_\nu=\mathbbm{1}$, whereas the 
matrix diagonalizing $f_R$ via 
$V_R^T \, f_R \, V_R = f_R^{\rm diag}$ is 
$$
V_R = \left(
\bad
\sqrt{\frac 12} & \sqrt{\frac 12} & 0 \\
-\frac 12 & \frac 12 & -\sqrt{\frac 12} \\
-\frac 12 & \frac 12 & \sqrt{\frac 12}  
\ea
\right) P_R\,, 
$$
with 
$P_R = {\rm diag}(i , 1, e^{-i \phi/2})$.
The 
eigenvalues of $f_R$ are 
$\epsilon \, (1 - \sqrt{2} \, B) $,  
$\epsilon \, (1 + \sqrt{2} \, B)$ and $e^{i \phi}$. Since we 
started in a basis in which $f_R = f_L$ is not diagonal, we have to use a 
modified parameterization for $Y_D$, which is obtained from 
Eq.~(\ref{eq:paraII}) by multiplying it on the right by $V_R$. The Dirac-type 
Yukawa coupling matrix is then found to be  
\bea
\D  Y_D =  
i (\sqrt{v_R}/v_u) \, V_\nu^\ast \, 
\sqrt{X_\nu^{\rm diag}} \, R \, \sqrt{f_R^{\rm diag}} \, V_R  
=   i (\sqrt{v_R \, v_L}/v_u) \, {\rm diag}(0,0,\sqrt{\eta}) \, 
R \, \sqrt{f_R^{\rm diag}} \, V_R  \\ \D 
= \left( 
\bad 
0 & 0 & 0 \\
0 & 0 & 0 \\
(Y_D)_{31} & (Y_D)_{32} & (Y_D)_{33} 
\ea
\right)\,,
\eea
where the non-zero entries are 
\begin{eqnarray} \nonumber 
(Y_D)_{31} & = &  \, \frac{\sqrt{v_L \, \eta}}{2 \, v_u}
 \left(\sqrt{2} \, \sqrt{M_1} \, \sin \omega_{13} 
+ \cos  \omega_{13}  \, (\sqrt{M_3} \, \cos \omega_{23} - 
 \sqrt{M_2} \, \sin \omega_{23} )\right)\,,\\ \nonumber 
(Y_D)_{32} & = & -i \, \frac{\sqrt{v_L \, \eta}}{2 \, v_u}
\left(\sqrt{2} \, \sqrt{M_1} \, \sin \omega_{13} 
- \cos  \omega_{13}  \, (\sqrt{M_3} \, \cos \omega_{23} - 
 \sqrt{M_2} \, \sin \omega_{23} )\right) \,,\\ \nonumber 
(Y_D)_{33} & = & i \, \frac{\sqrt{v_L \, \eta}}{\sqrt{2} \, v_u} 
e^{-i \, \phi/2} 
\cos  \omega_{13}  \, (\sqrt{M_3} \, \cos \omega_{23} + 
 \sqrt{M_2} \, \sin \omega_{23} )\,.
\end{eqnarray}
Note that in all three examples we have considered so far, $Y_D$ does not 
depend on $\omega_{12}$. This is related to the fact that in all these  
examples the matrix $X_\nu^{\rm diag}$ has only one (namely, third) 
non-vanishing diagonal entry.

The result for LFV in the present example is that the branching ratios of 
the decays $\ell_i \ra \ell_j \gamma$ 
depend only on $f_L \, f_L^\dagger$, namely,  
$(f_L \, f_L^\dagger)_{12} = (f_L \, f_L^\dagger)_{13} 
= 2 \, \epsilon^2 \, B $ and  
$(f_L \, f_L^\dagger)_{23} = 
-\frac 12 [1 - \epsilon^2 \, (1 + 2 \, B^2) ]$. 
As in the previous two examples, 
$\tau \ra e \gamma$ is too rare to be observable. The ratio 
BR$(\mu \ra e \gamma)$/BR$(\tau \ra \mu \gamma)$ is approximately  
$(2 \, \epsilon^2 \, B/\frac12)^2/0.174 \simeq 10^{-3}$.

\begin{figure}[t]
\begin{center}
\epsfig{file=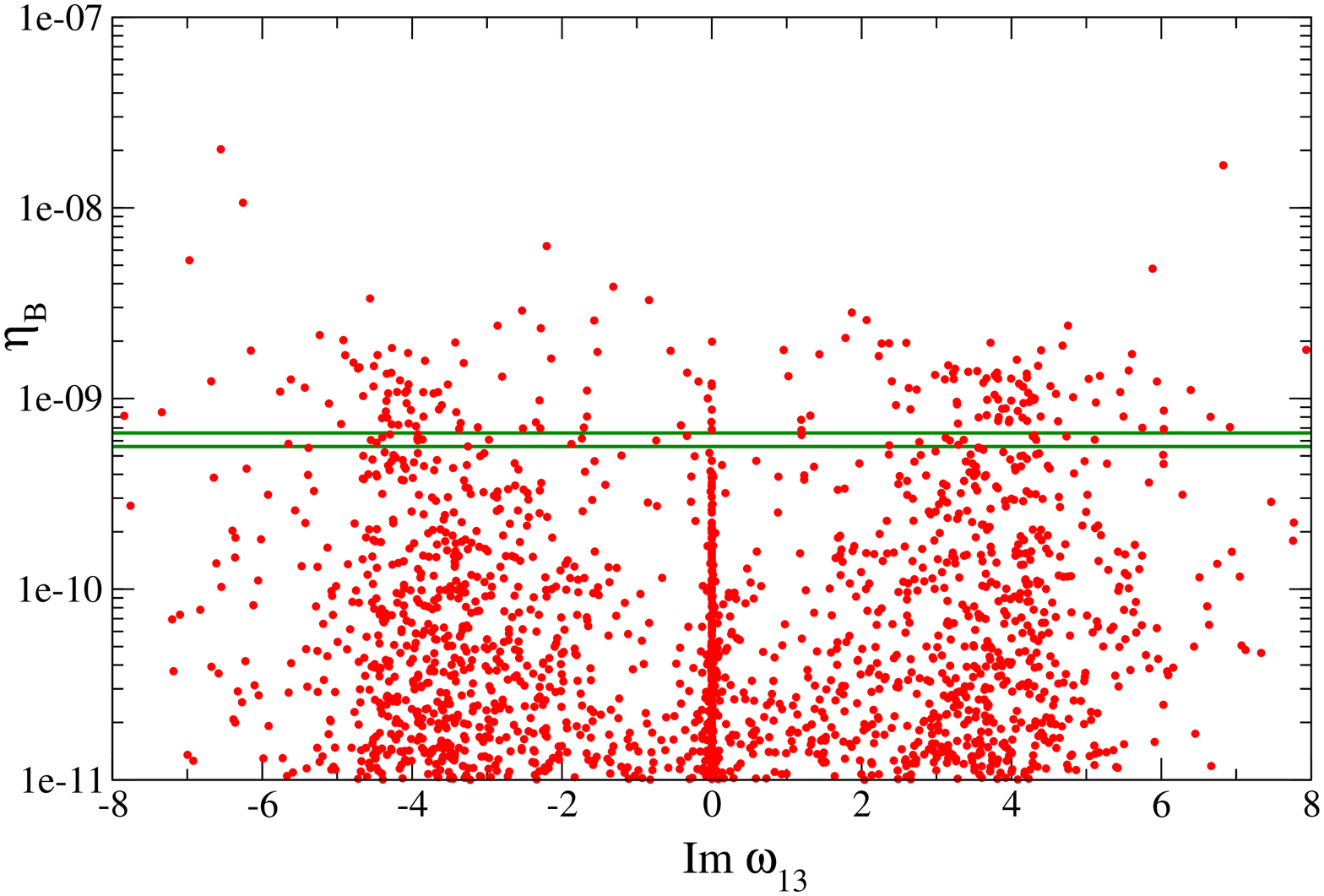,width=12cm,height=8cm}
\caption{\label{fig:ex4}Scatter plot for the baryon 
asymmetry $\eta_B$ 
against the imaginary part of $\omega_{13}$ for the 
third example of Section~\ref{sec:third}. The observed value of 
$\eta_B$ corresponds to 
the region between the horizontal lines. }
\end{center}

\end{figure}

\begin{figure}[t]
\begin{center}
\epsfig{file=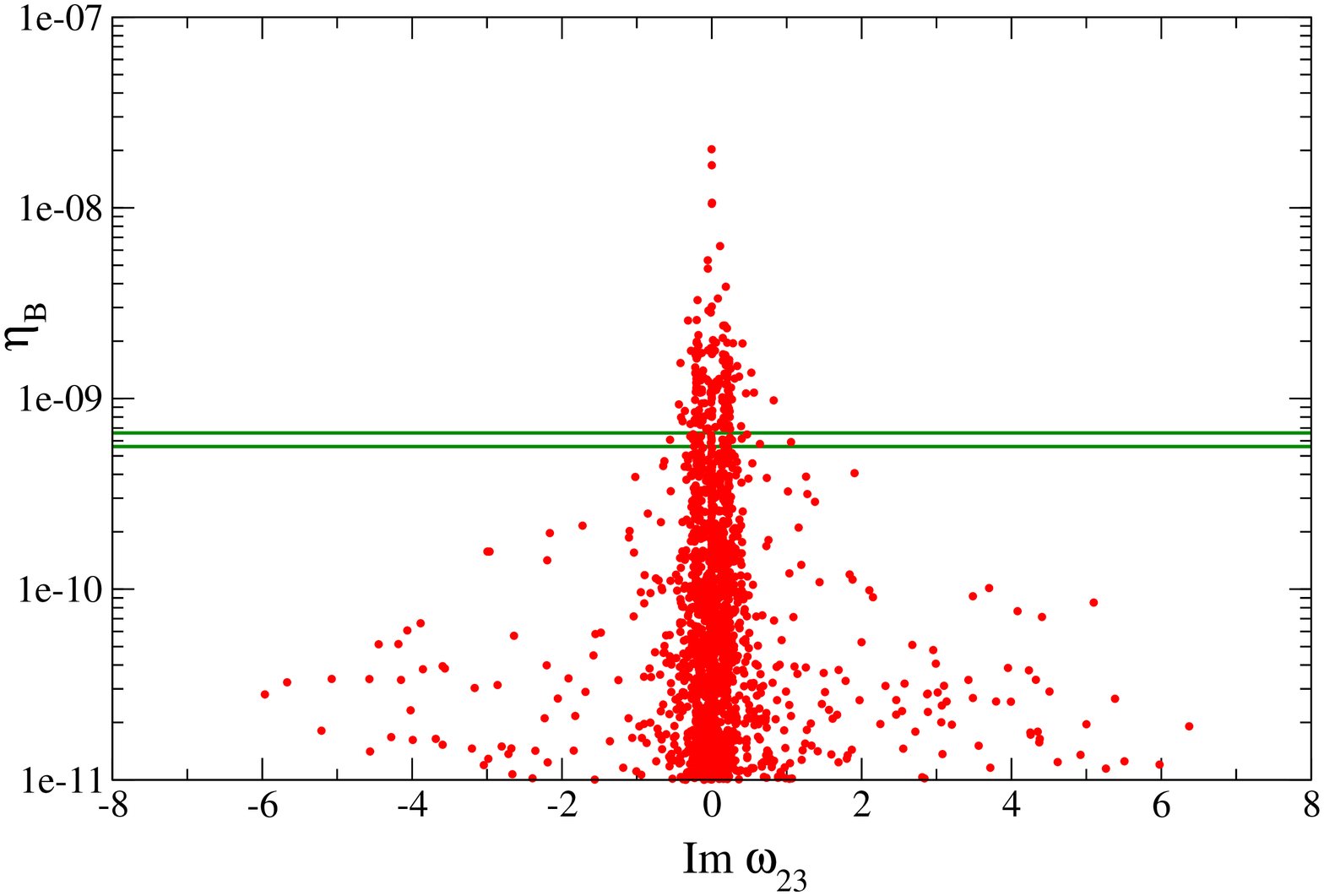,width=12cm,height=8cm}
\caption{\label{fig:ex5}Scatter plot for the baryon 
asymmetry $\eta_B$ 
against the imaginary part of $\omega_{23}$ for the 
third example of Section~\ref{sec:third}. The observed value of 
$\eta_B$ corresponds to 
the region between the horizontal lines. }
\end{center}
\end{figure}

Turning to leptogenesis, only 
$(\varepsilon_1^\tau)_N$ and 
$(\varepsilon_1^\tau)_\Delta$ are non-zero. 
The corresponding expressions
are rather lengthy and we do not give them here.  
Figs.~\ref{fig:ex4} and \ref{fig:ex5} show scatter 
plots of the baryon asymmetry against the imaginary 
parts of $\omega_{13}$ 
and $\omega_{23}$ for fixed values of the LFV 
branching ratios.  
For definiteness, we have chosen again the SUSY 
parameters $\tan \beta = 10$, 
$m_0 = 1450$ GeV, $m_{1/2} = 300$ GeV and $A_0 = 0$, which gives 
BR$(\mu \ra e \gamma) = 3.0 \cdot 10^{-12}$, 
BR$(\tau \ra e \gamma) = 5.4 \cdot 10^{-13}$ and 
BR$(\tau \ra \mu \gamma) = 3.1 \cdot 10^{-9}$.

\section{\label{sec:concl}Summary and conclusions}
We have considered lepton flavor violation and leptogenesis in the
case of type I + II seesaw, when the exchanges of both heavy Majorana
neutrinos and $SU(2)_L$-triplet Higgs bosons contribute to the mass matrix
of light neutrinos. We have proposed a parameterization of the Dirac-type
neutrino Yukawa coupling matrix $Y_D$ in this framework, which generalizes
the Casas-Ibarra parameterization suggested for type I seesaw. 
Our parameterization automatically takes into account the type I + II
seesaw formula and, like the Casas-Ibarra one, involves an arbitrary
complex orthogonal matrix $R$. This matrix depends in general on six real
parameters and can be parameterized in terms of three complex angles.
We have given simple examples illustrating the usefulness of the proposed
parameterization. In particular, we have considered LFV decays $\ell_i \ra
\ell_j \gamma$ and leptogenesis in the case when the type I and type II
contributions to both the light neutrino mass matrix $m_\nu$ and the
slepton mass matrix $\tilde{m}_L^2$ governing the LFV decays are of the
same order. We considered two examples leading to the tri-bimaximal leptonic
mixing and an example based on a relatively small but phenomenologically viable
deviation from bimaximal mixing. In all the examples we have studied we
found that the matrix $Y_D$ depends only on two out of the three complex 
angles parameterizing the matrix $R$, which is related to the fact that 
the matrix $X_\nu\equiv m_\nu - f_L \, v_L$ had 
only one non-zero eigenvalue.

In each of the three examples that we considered, we have found that 
the decays $\mu\ra e \gamma$ and $\tau\ra e\gamma$ are governed by the
same quantity, and the corresponding branching ratios are related by
BR$(\tau\ra e\gamma)\simeq 0.178\,$BR$(\mu\ra e\gamma)$, which is a
consequence of the approximate $\mu$-$\tau$ symmetry of the involved mass 
matrices.

In the first two examples based on tri-bimaximal leptonic mixing we found 
that leptogenesis is essentially governed by one of the three complex 
angles parameterizing the matrix $R$. This can be traced back to the facts  
that the masses of heavy Majorana neutrinos were assumed to be hierarchical 
and that the loops with the triplet exchange gave no contribution to 
lepton asymmetry in these examples. 

To conclude, we proposed a new parameterization of the Dirac-type neutrino
Yukawa coupling matrix $Y_D$ which is the most general one satisfying
the combined type I + II seesaw formula. It expresses the matrix $Y_D$ 
through both low energy and high energy parameters and can be useful for 
studies of lepton flavor violation and leptogenesis in the type I + II 
seesaw framework.

\vspace{0.3cm}
\begin{center}
{\bf Acknowledgments}
\end{center}
We thank S.~Antusch, S. Davidson. M.~Frigerio, E.~Nardi 
and Y.~Nir for useful discussions. 
This work was supported in part by the 
Deutsche Forschungsgemeinschaft 
in the Transregio 27 ``Neutrinos and beyond -- weakly interacting 
particles in physics, astrophysics and cosmology'' (W.R.).

\clearpage



\begin{thebibliography}{99}


\bibitem{I}
P.~Minkowski, 
Phys.\ Lett.\ B {\bf 67}, 421 (1977); 
T.~Yanagida, \emph{Horizontal gauge symmetry and masses of neutrinos}, in
  \emph{Proceedings of the Workshop on The Unified Theory and the Baryon Number
  in the Universe} (O.~Sawada and A.~Sugamoto, eds.), KEK, Tsukuba, Japan,
  1979, p.~95;
S.~L. Glashow, \emph{The future of elementary particle physics}, in
  \emph{Proceedings of the 1979 Carg{\`e}se Summer Institute on Quarks and
  Leptons} (M.~L{\'e}vy, J.-L. Basdevant, D.~Speiser, J.~Weyers, R.~Gastmans,
  and M.~Jacob, eds.), Plenum Press, New York, 1980, p.~687;
M.~Gell-Mann, P.~Ramond, and R.~Slansky, \emph{Complex spinors and unified
  theories}, in \emph{Supergravity} (P.~van Nieuwenhuizen and D.~Z. Freedman,
  eds.), North Holland, Amsterdam, 1979, p.~315;
R.~N. Mohapatra and G.~Senjanovi{\'c}, 
Phys.\ Rev.\ Lett.\ \textbf{44}, 912 (1980). 

\bibitem{II}
M.~Magg and C.~Wetterich,
Phys.\ Lett.\ B {\bf 94}, 61 (1980); 
R.~N.~Mohapatra and G.~Senjanovic,
%
Phys.\ Rev.\ D {\bf 23}, 165 (1981); 
G.~Lazarides, Q.~Shafi and C.~Wetterich,
Nucl.\ Phys.\ B {\bf 181}, 287 (1981); 
J.~Schechter and J.~W.~F.~Valle,
Phys.\ Rev.\ D {\bf 22}, 2227 (1980).

\bibitem{reviews}R.~N.~Mohapatra {\it et al.}, 
Rept.\ Prog.\ Phys.\  {\bf 70}, 1757 (2007); 
R.~N.~Mohapatra and A.~Y.~Smirnov,
   Ann.\ Rev.\ Nucl.\ Part.\ Sci.\  {\bf 56}, 569 (2006); 
A.~Strumia and F.~Vissani,
  hep-ph/0606054.




\bibitem{leptog} M.~Fukugita and T.~Yanagida,
  Phys.\ Lett.\  B {\bf 174}, 45 (1986).



\bibitem{D}For a recent review, see 
S.~Davidson, E.~Nardi and Y.~Nir,
  arXiv:0802.2962 [hep-ph].



\bibitem{Ilepto} 
M.~A.~Luty,
  Phys.\ Rev.\  D {\bf 45}, 455 (1992); 
M.~Flanz, E.~A.~Paschos and U.~Sarkar,
  Phys.\ Lett.\  B {\bf 345}, 248 (1995)
  [Erratum-ibid.\  B {\bf 382}, 447 (1996)]; 
  M.~Flanz, E.~A.~Paschos, U.~Sarkar and J.~Weiss,
  Phys.\ Lett.\  B {\bf 389}, 693 (1996); 
  A.~Pilaftsis,
  Phys.\ Rev.\  D {\bf 56}, 5431 (1997); 
  W.~Buchm\"uller and M.~Pl\"umacher,
  Phys.\ Lett.\  B {\bf 431}, 354 (1998).




\bibitem{CRV} L.~Covi, E.~Roulet and F.~Vissani,
  Phys.\ Lett.\  B {\bf 384}, 169 (1996). 


\bibitem{leptoIIpure} E.~Ma and U.~Sarkar,
  Phys.\ Rev.\ Lett.\  {\bf 80}, 5716 (1998); 
T.~Hambye, E.~Ma and U.~Sarkar,
  Nucl.\ Phys.\  B {\bf 602}, 23 (2001). 

\bibitem{HamSen} 
T.~Hambye and G.~Senjanovi{\'c},
  Phys.\ Lett.\  B {\bf 582}, 73 (2004); 
 S.~Antusch and S.~F.~King,
  Phys.\ Lett.\  B {\bf 597}, 199 (2004); 
S.~Antusch,
  Phys.\ Rev.\  D {\bf 76}, 023512 (2007).

\bibitem{UO}
 P.~J.~O'Donnell and U.~Sarkar,
  Phys.\ Rev.\  D {\bf 49}, 2118 (1994). 

\bibitem{JPR0}A.~S.~Joshipura, E.~A.~Paschos and W.~Rodejohann,
  Nucl.\ Phys.\  B {\bf 611}, 227 (2001). 


\bibitem{JPR1} A.~S.~Joshipura, E.~A.~Paschos and W.~Rodejohann,
  JHEP {\bf 0108}, 029 (2001); 
W.~Rodejohann,
  Phys.\ Lett.\  B {\bf 542}, 100 (2002). 



\bibitem{ILFV}F.~Borzumati and A.~Masiero,
  Phys.\ Rev.\ Lett.\  {\bf 57}, 961 (1986); 
J.~Hisano, T.~Moroi, K.~Tobe and M.~Yamaguchi,
  Phys.\ Rev.\ D {\bf 53}, 2442 (1996). 

\bibitem{IILFV}
E.~J.~Chun, K.~Y.~Lee and S.~C.~Park,
  Phys.\ Lett.\  B {\bf 566}, 142 (2003); 
M.~Kakizaki, Y.~Ogura and F.~Shima,
  Phys.\ Lett.\  B {\bf 566}, 210 (2003). 
 

\bibitem{Rossi}
A.~Rossi,
  Phys.\ Rev.\  D {\bf 66}, 075003 (2002). 


\bibitem{LFVIIana}E.~J.~Chun, A.~Masiero, A.~Rossi and S.~K.~Vempati,
  Phys.\ Lett.\  B {\bf 622}, 112 (2005); 
F.~R.~Joaquim and A.~Rossi,
  Phys.\ Rev.\ Lett.\  {\bf 97}, 181801 (2006); 
  Nucl.\ Phys.\  B {\bf 765}, 71 (2007). 


\bibitem{CI} J.~A.~Casas and A.~Ibarra,
  Nucl.\ Phys.\  B {\bf 618}, 171 (2001).

\bibitem{LFVIana}For instance, 
 S.~Lavignac, I.~Masina and C.~A.~Savoy,
  Phys.\ Lett.\  B {\bf 520}, 269 (2001); 
J.~R.~Ellis, J.~Hisano, S.~Lola and M.~Raidal,
  Nucl.\ Phys.\  B {\bf 621}, 208 (2002); 
 A.~Kageyama, S.~Kaneko, N.~Shimoyama and M.~Tanimoto,
  Phys.\ Lett.\  B {\bf 527}, 206 (2002); 
S.~Lavignac, I.~Masina and C.~A.~Savoy,
  Nucl.\ Phys.\  B {\bf 633}, 139 (2002); 
J.~R.~Ellis, J.~Hisano, M.~Raidal and Y.~Shimizu,
  Phys.\ Rev.\  D {\bf 66}, 115013 (2002); 
 F.~Deppisch, H.~P\"as, A.~Redelbach, R.~R\"uckl and Y.~Shimizu,
  Eur.\ Phys.\ J.\  C {\bf 28}, 365 (2003); 
A.~Masiero, S.~K.~Vempati and O.~Vives,
  Nucl.\ Phys.\  B {\bf 649}, 189 (2003); 
S.~Pascoli, S.~T.~Petcov and C.~E.~Yaguna,
  Phys.\ Lett.\  B {\bf 564}, 241 (2003); 
S.~Pascoli, S.~T.~Petcov and W.~Rodejohann,
  Phys.\ Rev.\  D {\bf 68}, 093007 (2003); 
S.~Kanemura, K.~Matsuda, T.~Ota, T.~Shindou, E.~Takasugi and K.~Tsumura,
  Phys.\ Rev.\  D {\bf 72}, 093004 (2005); 
P.~Di Bari,
  Nucl.\ Phys.\  B {\bf 727}, 318 (2005); 
S.~T.~Petcov, W.~Rodejohann, T.~Shindou and Y.~Takanishi,
  Nucl.\ Phys.\  B {\bf 739}, 208 (2006); 
 F.~Deppisch, H.~P\"as, A.~Redelbach and R.~R\"uckl,
  Phys.\ Rev.\  D {\bf 73}, 033004 (2006); 
 S.~T.~Petcov and T.~Shindou,
  Phys.\ Rev.\  D {\bf 74}, 073006 (2006); 
S.~Antusch, E.~Arganda, M.~J.~Herrero and A.~M.~Teixeira,
  JHEP {\bf 0611}, 090 (2006); 
 G.~C.~Branco, A.~J.~Buras, S.~Jager, S.~Uhlig and A.~Weiler,
   JHEP {\bf 0709}, 004 (2007); 
J.~A.~Casas, A.~Ibarra and F.~Jimenez-Alburquerque,
  JHEP {\bf 0704}, 064 (2007). 


\bibitem{PPTS}
S.~T.~Petcov, S.~Profumo, Y.~Takanishi and C.~E.~Yaguna,
  Nucl.\ Phys.\  B {\bf 676}, 453 (2004). 


\bibitem{newdata}
M.~C.~Gonzalez-Garcia and M.~Maltoni,
  arXiv:0704.1800 [hep-ph].




\bibitem{AF1}
  E.~K.~Akhmedov and M.~Frigerio,
  Phys.\ Rev.\ Lett.\  {\bf 96}, 061802 (2006).





\bibitem{Host}
  P.~Hosteins, S.~Lavignac and C.~A.~Savoy,
  Nucl.\ Phys.\  B {\bf 755}, 137 (2006). 

\bibitem{AF2}
  E.~K.~Akhmedov and M.~Frigerio,
  JHEP {\bf 0701}, 043 (2007). 

\bibitem{LR}
M.~Lindner and W.~Rodejohann,
   JHEP {\bf 0705}, 089 (2007).


\bibitem{PDG}W.~M.~Yao {\it et al.}  [Particle Data Group],
  J.\ Phys.\ G {\bf 33}, 1 (2006).



\bibitem{mueg_lim}
M.~L.~Brooks {\it et al.}  [MEGA Collaboration],
  Phys.\ Rev.\ Lett.\  {\bf 83}, 1521 (1999). 

\bibitem{teg_lim}B.~Aubert {\it et al.}  [BABAR Collaboration],
  Phys.\ Rev.\ Lett.\  {\bf 96}, 041801 (2006). 

\bibitem{tmg_lim}B.~Aubert {\it et al.}  [BABAR Collaboration],
  Phys.\ Rev.\ Lett.\  {\bf 95}, 041802 (2005). 


\bibitem{meg_fut}See the homepage of the MEG experiment, 
{\tt http://meg.web.psi.ch}. 


\bibitem{BR_fut}A.~G.~Akeroyd {\it et al.}, hep-ex/0406071. 








\bibitem{LFV_rev}
A.~Masiero, S.~K.~Vempati and O.~Vives,
  New J.\ Phys.\  {\bf 6}, 202 (2004). 








\bibitem{flavor_flav}
A.~Abada, S.~Davidson, F.~X.~Josse-Michaux, M.~Losada and A.~Riotto,
  JCAP {\bf 0604}, 004 (2006); 
 E.~Nardi, Y.~Nir, E.~Roulet and J.~Racker,
  JHEP {\bf 0601}, 164 (2006); 
see also 
R.~Barbieri, P.~Creminelli, A.~Strumia and N.~Tetradis,
  Nucl.\ Phys.\  B {\bf 575}, 61 (2000); 
recent overviews are given in 
 S.~Blanchet and P.~Di Bari,
  Nucl.\ Phys.\ Proc.\ Suppl.\  {\bf 168}, 372 (2007) [hep-ph/0702089]; 
S.~Davidson,
  arXiv:0705.1590 [hep-ph].


\bibitem{PB} S.~Blanchet and P.~Di Bari, 
  JCAP {\bf 0606}, 023 (2006); 
  JCAP {\bf 0703}, 018 (2007). 














\bibitem{tri}
P.~F.~Harrison, D.~H.~Perkins and W.~G.~Scott,
  Phys.\ Lett.\ B {\bf 530}, 167 (2002); 
  Phys.\ Lett.\ B {\bf 535}, 163 (2002); 
Z.~Z.~Xing,
  Phys.\ Lett.\ B {\bf 533}, 85 (2002); 
  X.~G.~He and A.~Zee,
  Phys.\ Lett.\ B {\bf 560}, 87 (2003); 
see also 
L.~Wolfenstein,
  Phys.\ Rev.\ D {\bf 18}, 958 (1978); 
Y.~Yamanaka, H.~Sugawara and S.~Pakvasa,
  Phys.\ Rev.\  D {\bf 25}, 1895 (1982)
  [Erratum-ibid.\  D {\bf 29}, 2135 (1984)]. 


\bibitem{bima}F.~Vissani, hep-ph/9708483; 
V.~D.~Barger, S.~Pakvasa, T.~J.~Weiler and K.~Whisnant, 
Phys.\ Lett.\ B {\bf 437}, 107 (1998); 
A.~J.~Baltz, A.~S.~Goldhaber and M.~Goldhaber, 
Phys.\ Rev.\ Lett.\  {\bf 81}, 5730 (1998); 
H.~Georgi and S.~L.~Glashow,
Phys.\ Rev.\ D {\bf 61}, 097301 (2000); 
I.~Stancu and D.~V.~Ahluwalia,
Phys.\ Lett.\ B {\bf 460}, 431 (1999). 




\bibitem{SPS}B.~C.~Allanach {\it et al.},
  Eur.\ Phys.\ J.\ C {\bf 25}, 113 (2002)
  [eConf {\bf C010630}, P125 (2001)]. 

\bibitem{WRII}W.~Rodejohann,
  Phys.\ Rev.\  D {\bf 70}, 073010 (2004). 







\end{thebibliography}
\end{document}